\documentclass[11pt]{amsart}
\usepackage{amscd}
\usepackage{amsmath}
\usepackage{graphicx}
\usepackage{amsfonts}
\usepackage{amssymb}
\begin{document}

\title[Entanglement detection beyond the CCNR criterion for infinite-dimensions]
{Entanglement detection beyond the CCNR criterion for infinite-dimensions}

\date{ \today}

\author{Yu Guo}
\address[Y. Guo]
{School of
Mathematics,
Taiyuan University of Technology,
 Taiyuan 030024,
  P. R. China;
Department of Mathematics, Shanxi Datong University, Datong 037009,
China;
Department of Physics and
Optoelectroincs, Taiyuan University of Technology, Taiyuan 030024,
China}
\email{guoyu3@yahoo.com.cn}

\author{Jinchuan Hou}
\address[J. Hou]{School of
Mathematics,
Taiyuan University of Technology,
 Taiyuan 030024,
  P. R. China}
\email{jinchuanhou@yahoo.com.cn, houjinchuan@tyut.edu.cn}




\maketitle

\begin{abstract}
In this paper, in terms of the relation between
the state and its reduced states, we obtain two inequalities
which are valid for all separable states in infinite-dimensional bipartite
quantum systems. One of them
provides an entanglement criterion which is strictly stronger than
the computable cross-norm or realignment (CCNR) criterion.
\end{abstract}


\section{Introduction}

Quantum entanglement has been subjected to intensive studies in
connection with quantum information theory and quantum communication
theory \cite{1}. One basic problem for quantum entanglement is
to find a proper criterion to determine whether a given state of a
composite system is entangled or not \cite{2,3,4,5,6,7,8,Hou6,Hou7,Hou8,Hou9}.
Although considerable progress has been achieved in this field, this problem
is not fully explored yet except for the case of $2\otimes 2$ and $2\otimes 3$
systems \cite{2,3,9}.

By definition, a bipartite state $\rho$ acting on a separable
complex Hilbert space $H=H_A\otimes H_B$ is called \emph{separable}
if and only if it can be written as
$$\rho=\sum\limits_ip_i\rho_i^A\otimes\rho_i^B,\quad \sum\limits_ip_i=1, \
p_i\geq0 \eqno{(1)}$$ or it is a limit of the states of the above
form under the trace norm topology \cite{10}, where $\rho_i^A$ and
$\rho_i^B$ are (pure) states on the subsystems associated to the
Hilbert spaces $H_A$ and $H_B$, respectively. A state that is not
separable is said to be \emph{entangled}. Particularly, if a state
can be represented in the form as in Eq.(1), it is called
\emph{countably separable} \cite{11}. Observing that, for
finite-dimensional systems, all separable states are finitely
separable. However, there do exist some separable
states which are
not countably separable in infinite-
dimensional systems
\cite{11}.

For finite-dimensional systems, a very elegant criterion for the
separability is the so-called \emph{computable cross norm} or
\emph{realignment} (CCNR) criterion proposed by Rudolph in \cite{12}
and Chen and Wu \cite{13}. The CCNR criterion states that if $\rho$
is a separable state on $H_A\otimes H_B$ with $\dim H_A\otimes
H_B<+\infty $, then the trace norm $\|\rho^R\|_{\rm Tr}$ of the
realignment matrix  $ \rho^R$ of $\rho$ is not greater than 1. By
exploring the relation between the state and its reduced states,
Zhang \emph{et al} \cite{14} investigated a criterion beyond the
CCNR criterion. It is showed in \cite{14} that a state acting on
$H_A\otimes H_B$ with $\dim H_A\otimes H_B<+\infty$ is separable
implies that
$$\|(\rho-\rho_A\otimes\rho_B)^{R}\|_{\rm Tr}\leq\sqrt{[1-{\rm
Tr}(\rho_A^2)][1-{\rm Tr}(\rho_B^2)]}\eqno{(2)}$$ and
$$\|(\rho-\rho_A\otimes\rho_B)^{T_B}\|_{\rm
Tr}\leq2\sqrt{[1-{\rm Tr}(\rho_A^2)][1-{\rm
Tr}(\rho_B^2)]}.\eqno{(3)}$$ Here, $C^R$ denotes the realignment
matrix of the block matrix $C=[C_{ij}]_{N_A\times N_A}$ with
$C_{ij}$s are $N_B\times N_B$ complex matrices, where $\dim H_A=N_A$
and $\dim H_B=N_B$. $\|\cdot\|_{\rm Tr}$ denotes the trace norm and
$C^{T_B}$ denotes the partial transposition of $C$ with respect to
the subsystem B. The inequality (2) provides a criterion which is
stronger than the CCNR criterion \cite{14} (namely, any entangled states
that detected by the CCNR criterion can be detected by the inequality (2)
and there exist some entangled states that can be detected by inequality (2)
while they can't be recognized by the CCNR criterion).

Very recently, we established the realignment operation and CCNR
criterion for infinite-dimensional bipartite systems \cite{15,16}.
It is showed in \cite{15} that $\|\rho^R\|_{\rm Tr}\leq1$ whenever
$\rho$ is a separable state acting on $H_A\otimes H_B$ with $\dim
H_A\otimes H_B\leq+\infty$. The aim of this paper is to establish
the analogous inequalities as (2) and (3) for infinite-dimensional
case. In addition, as one might expect, we show that the obtained
criterion is stronger than the CCNR criterion proposed in \cite{15},
and furthermore, it can detect some PPT entangled states (i.e, the
entangled states with positive partial transposition) which can not
be detected by the CCNR criterion.
It should be pointed out that the corresponding
inequalities for infinite-dimensional case can not be derived
straightforwardly from that of the
finite-dimensional case. The situations grow more complicated in
the case of infinite-dimensional case.

In detail, our paper is organized as follows. In Sec.II we propose
some properties of the reduced density operators for both finite- and
infinite-dimensional bipartite systems. We show that the reduced
states stand close to each other whenever the composite states are
closed to each other. Then in Sec.III we propose a practical criterion
based on the relation $\rho-\rho_A\otimes \rho_B$. The obtained criterion is strictly
stronger than the CCNR criterion.  Sec.IV is a short conclusion.

Throughout the paper, we use the bra-ket notations.
$\langle\cdot|\cdot\rangle$ stands for the inner product in the
given Hilbert spaces. The set of all (bounded linear) operators on a
Hilbert space $H$ is denoted by $\mathcal{B}(H)$, the set of all
trace class operators on $H$ is denoted by $\mathcal{T}(H)$ and the
space consisting of all Schattern-p class operators on $H$ is
denoted by $\mathcal{C}_p(H)$. $A\in\mathcal{B}(H)$ is self-adjoint
if $A^\dagger=A$ ($A^\dagger$ stands for the adjoint operator of
$A$); $A$ is said to be positive, denoted by $A\geq0$, if
$A^\dagger=A$ and $\langle\psi|A|\psi\rangle\geq0$ for all
$|\psi\rangle\in H$. $A^T$ stands for the transposition of the
operator $A$. By $\mathcal{S}(H_A)$, $\mathcal{S}(H_B)$ and
$\mathcal{S}(H_A\otimes H_B)$ we denote the sets of all states
acting on $H_A$, $H_B$ and $H_A\otimes H_B$, respectively. By
$\mathcal{S}_{sep}(H_A\otimes H_B)$ we denote the set of all
separable states in $\mathcal{S}(H_A\otimes H_B)$. \if A state
$\rho$ is called a pure state if ${\rm Tr}(\rho^2)=1$ and if ${\rm
Tr}(\rho^2)<1$ it is called mixed as usual. We also call a unit
vector $|\psi\rangle\in H_A\otimes H_B$ pure state which is
corresponding to the density operator
$\rho=|\psi\rangle\langle\psi|$.\fi  We fix in the `local state
spaces' $H_A$ and $H_B$ orthonormal bases
$\{|m\rangle\}_{m=1}^{N_A}$ and $\{|\mu\rangle\}_{\mu=1}^{N_B}$,
respectively, where $\dim H_A=N_A$ and $\dim H_B=N_B$ ($N_{A/ B}$
may be $+\infty$)  (note that we use Latin indices for the subsystem
$\rm{A}$ and the Greek indices for the subsystem $\rm{B}$). \if For
a state $\rho\in\mathcal{S}$, $\rho_A={\rm Tr}_B(\rho)$ and
$\rho_B={\rm Tr}_A(\rho)$ are the reduced states of $\rho$ with
respect to the subsystems ${\rm B}$ and ${\rm A}$, respectively. It
is easy to see that if $\rho$ is separable and
$\rho=\sum\limits_ip_i\rho_i^A\otimes\rho_i^B$ with $p_i\geq0$,
$\sum\limits_ip_i=1$, $\rho_i^A\in\mathcal{S}^A$ and
$\rho_i^B\in\mathcal{S}^B$, then $\rho_A=\sum\limits_ip_i\rho_i^A$
and $\rho_B=\sum\limits_ip_i\rho_i^B$.\fi The partial transposition
of $\rho\in\mathcal{S}(H_A\otimes H_B)$ with respect to the subsystem ${\rm B}$
(resp. ${\rm A}$) is denoted by $\rho^{T_B}$ (resp. $\rho^{T_A}$),
that is, $\rho^{T_B}=(I_A\otimes{\bf T})\rho$ (resp.
$\rho^{T_A}=({\bf T}\otimes I_B)\rho$), where ${\bf T}$ is the map
of taking transpose, ${\bf T}C=C^T$,  with respect to a given
orthonormal basis.

\section{The reduced density operators}

To describe subsystems of a composite system, one needs the reduced density
operator. It is so useful as to be virtually indispensable in the analysis of composite
systems \cite{1}. In this section, we discuss some properties about the reduced density operators.

Let $H_A$ and $H_B$ be complex Hilbert spaces with $\dim H_A\otimes H_B=+\infty$,
$\rho=|\psi\rangle\langle\psi|\in\mathcal{S}(H_A\otimes H_B)$ be a
pure state. We write
$|\psi\rangle=\sum\limits_{m,\mu}d_{m\mu}|m\rangle|\mu\rangle$. It
is clear that $D_{\psi}=(d_{m\mu})$ can be regarded as an operator
from $H_B$ into $H_A$ and it is a Hilbert-Schmidt class operator
with the Hilbert-Schmidt norm $\|D_{\psi}\|_2=\||\psi\rangle\|$. Under the given bases, we have
$$\begin{array}{rcl}\rho_A&=&{\rm Tr}_B(\rho)=(I_A\otimes {\rm\bf Tr})\rho\\
&=&(I_A\otimes {\rm\bf Tr})
(\sum\limits_{m,\mu,n,\nu}d_{m\mu}\bar{d_{n\nu}}|m\rangle\langle n|
\otimes|\mu\rangle\langle\nu|)\\
&=&\sum\limits_{m,\mu,n,\nu}d_{m\mu}\bar{d_{n\nu}}
{\rm Tr}(|\mu\rangle\langle\nu|)|m\rangle\langle n|\\
&=&\sum\limits_{m,n,\mu}d_{m\mu}\bar{d_{n\mu}}|m\rangle\langle n|\\
&=&\sum\limits_{m,n}(\sum\limits_\mu d_{m\mu}\bar{d_{n\mu}})|m\rangle\langle n|=DD^\dag.\end{array}$$
Similarly, ${\rm Tr}_A(\rho)=( {\rm\bf Tr}\otimes
I_B)\rho=\rho_B=D^\dag D$.  For any
mixed state $\rho\in\mathcal{S}(H_A\otimes H_B)$,  let
$$\rho=\sum\limits_ip_i|\psi_i\rangle\langle\psi_i|,
\quad |\psi_i\rangle\in H_A\otimes H_B, \ p_i>0, \
\sum\limits_ip_i=1,$$ be the spectral decomposition. Write
$|\psi_i\rangle=\sum\limits_{m,\mu}d_{m\mu}^{(i)}|m\rangle|\mu\rangle$
and $D_i=(d_{m\mu}^{(i)})$. It turns out that
$$\rho_A=\sum\limits_ip_iD_iD_i^\dag,
\quad\rho_B=\sum\limits_ip_iD_i^\dag
D_i.\eqno{(4)}$$ That is, $\rho_A=\sum\limits_ip_iD_iD_i^\dag$
and $\rho_B=\sum\limits_ip_iD_i^\dag D_i$ are reduced density
operators [for the finite-dimensional case,
the discussion above is obvious (also see in \cite{17})].

If $\rho$, $\varrho\in{\mathcal S}(H_A\otimes H_B)$, and $\rho$ stands
close to $\varrho$, then, what about the distance between $\rho_{A/B}$ and
$\varrho_{A/B}$? In fact, we have\\

{\bf Proposition 1} \quad Let $H_A$ and $H_B$ be complex separable Hilbert spaces with $\dim H_A\otimes H_B\leq+\infty$,
$\rho$, $\rho_k\in{\mathcal S}(H_A\otimes H_B)$, $k=1$, 2, $\dots$ and
$\lim_k\rho_k=\rho$ in trace norm. Then
$$\lim\limits_{n\rightarrow\infty}\rho_{A(k)}
=\rho_A \quad \rm{and} \quad
\lim\limits_{n\rightarrow\infty}\rho_{B(k)}=\rho_B,\eqno{(5)}$$
in trace norm, where $\rho_{A(k)}={\rm Tr}_B(\rho_k)$ and
$\rho_{B(k)}={\rm Tr}_A(\rho_k)$.\\

{\bf Proof} \quad Take orthonormal bases $\{|m\rangle\}_{m=1}^{N_A}$ and
$\{|\mu\rangle\}_{\mu=1}^{N_B}$ of $H_A$ and $H_B$, respectively.
With respect to theses bases, we can write $\rho_k$ and $\rho$ in the
matrix form $\rho_k=(\sigma^{(k)}_{mn})$ and $\rho=(\sigma_{mn})$,
where $\sigma^{(k)}_{mn}, \sigma_{mn}\in{\mathcal T}(H_B)$. Then
$\rho_{A(k)}=({\rm Tr}(\sigma^{(k)}_{mn}))$ and $\rho_A=({\rm
Tr}(\sigma _{mn}))$. Since $\rho_k\rightarrow \rho$ as
$k\rightarrow\infty$ under the trace norm topology, we have
$\sigma^{(k)}_{mn}\rightarrow \rho_{mn}$ as $k\rightarrow\infty$
under the trace norm topology for each $(m,n)$-entry. Hence ${\rm
Tr}(\sigma^{(k)}_{mn})\rightarrow {\rm Tr}(\sigma _{mn})$ for any
$m,n$, that is, $\rho_{A(k)}$ converges to $\rho_A$ entry-wise. Note
that ${\mathcal T}(H)$ is the dual space of ${\mathcal B}_0(H)$, here
${\mathcal B}_0(H)$ denotes the
Banach space of all compact operators on $H$. It follows that,
$\rho_{A(k)}$ converges to $\rho_A$ under the weak star topology
$\sigma ({\mathcal T}(H),{\mathcal B}_0(H))$. It is known from
\cite{18} that the weak-star topology coincided with the trace norm
topology on ${\mathcal S}(H)$. Therefore, we conclude that
$\rho_{A(k)}\rightarrow \rho_A$ as $k\rightarrow\infty$ under the
trace norm topology.

Similarly, one can show that $\rho_k\rightarrow \rho$ as
$k\rightarrow\infty$ implies that $\rho_{B(k)}\rightarrow \rho_B$ as
$k\rightarrow\infty$. $\square$\\

This proposition also implies that the trace operation is completely
bounded under the trace norm topology on the set of all states.

\section{Detecting entanglement by inequalities induced from the CCNR criterion}

The main result of this section is the following.

{\bf Theorem 1} \ \ \ Let $H_A$ and $H_B$ be complex separable Hilbert spaces with $\dim H_A\otimes H_B=+\infty$,
$\rho\in{\mathcal S}_{sep}(H_A\otimes H_B)$.  Then
$$\|(\rho-\rho_A\otimes\rho_B)^{R}\|_{\rm Tr}\leq\sqrt{[1-{\rm
Tr}(\rho_A^2)][1-{\rm Tr}(\rho_B^2)]}\eqno{(6)}$$ and
$$\|(\rho-\rho_A\otimes\rho_B)^{T_B}\|_{\rm
Tr}\leq2\sqrt{[1-{\rm Tr}(\rho_A^2)][1-{\rm
Tr}(\rho_B^2)]},\eqno{(7)}$$ where $\rho_A={\rm Tr}_B(\rho)$,
$\rho_B={\rm Tr}_A(\rho)$, and $\rho^R$ stands for the
realignment operator of $\rho$.

There are three equivalent definitions of the realignment operator
of an  operator in $\mathcal{C}_2(H_A\otimes H_B)$ \cite{15},
one of them is the following:

{\bf Lemma 1} (Guo \emph{et al}. \cite{15}) \ \ \ Let $H_A$ and $H_B$ be complex Hilbert spaces with $\dim H_A\otimes
H_B=+\infty$ and let $C\in{\mathcal C}_2(H_A\otimes H_B)$ be a Hilbert-Schmidt operator with
$C=\sum\limits_kA_k\otimes B_k$, where
$A_k=\sum\limits_{m,n}a_{mn}^{(k)}|m\rangle\langle n|\in{\mathcal
C}_2(H_A)$,
$B_k=\sum\limits_{\mu,\nu}b_{\mu\nu}^{(k)}|\mu\rangle\langle
\nu|\in{\mathcal C}_2(H_B)$ and the series converges in
Hilbert-Schmidt norm. Then
$$C^R=\sum\limits_k|A_k\rangle\langle B_k|,\eqno{(8)}$$
where the series converges in Hilbert-Schmidt norm,
$|A_k\rangle=\sum\limits_{m,n}a_{mn}^{(k)}|m\rangle|n\rangle$,
$|B_k\rangle=\sum\limits_{\mu,\nu}b_{\mu\nu}^{(k)}|\mu\rangle|\nu\rangle$,
$\langle B_k|$ denotes the transposition of $|B_k\rangle$.

In order to prove Theorem 1, some more lemmas are needed. The
following lemma is well known for mathematicians and we include a
proof of it here for readers' convenience.

{\bf Lemma 2} \quad Let $H_A$ and $H_B$ be complex separable Hilbert spaces with $\dim H_A\otimes H_B=+\infty$,
$A\in\mathcal{C}_p(H_A)$, $B\in\mathcal{C}_p(H_B)$ and $1\leq
p<+\infty$. Then $A\otimes B\in\mathcal{C}_p(H_A\otimes H_B)$,
and further more, $$\|A\otimes B\|_p=\|A\|_p\|B\|_p.$$

{\bf Proof} \quad Let $A=U_1D_1V_1$ and $B=U_2D_2V_2$ be the singular
value decomposition of $A$ and $B$, respectively, where $D_1={\rm
diag}(\lambda_1$, $\lambda_2$, $\cdots$, $\lambda_n$, $\cdots)$ and
$D_2={\rm diag}(\lambda_1^\prime$, $\lambda_2^\prime$, $\cdots$,
$\lambda_n^\prime$, $\cdots)$ with
$\lambda_1\geq\lambda_2\geq\cdots\geq\lambda_n\geq\cdots$ and
$\lambda_1^\prime\geq\lambda_2^\prime\geq\cdots\geq\lambda_n^\prime\geq\cdots$.
It follows that
$$\|A\|_p=(\sum\limits_i\lambda_i^p)^{\frac{1}{p}}\ {\rm and} \ \|B\|_p
=(\sum\limits_i\lambda_i^{\prime p})^{\frac{1}{p}}.$$ Write
$U_1\otimes U_2=U$, $D_1\otimes D_2=D$ and $V_1\otimes V_2=V$. Then
we have $A\otimes B=(U_1D_1V_1)\otimes(U_2D_2V_2)=(U_1\otimes
U_2)(D_1\otimes D_2)(V_1\otimes V_2)=UDV$. Since    $D$ is a
diagonal operator with diagonal entries
$\{\lambda_i\lambda_j^\prime\}$, one sees that
$$\begin{array}{rl}&\|A\otimes B\|_p\\
=&(\sum\limits_{i,j}\lambda_i^p\lambda_j^{\prime p})^{\frac{1}{p}}\\
=&[\sum\limits_i\lambda_i^p(\sum\limits_j\lambda_j^{\prime p})]^{\frac{1}{p}}\\
=&(\sum\limits_i\lambda_i^p)^{\frac{1}{p}}
(\sum\limits_j\lambda_j^{\prime p})^{\frac{1}{p}}\\
=&\|A\|_p\|B\|_p,\end{array}$$ as desired.$\Box$

{\bf Lemma 3} \quad Let $H_A$ and $H_B$ be complex separable Hilbert spaces with $\dim H_A\otimes H_B=+\infty$ and
$\{\rho_k\}$ be a sequence in ${\mathcal S}_{sep}(H_A\otimes H_B)$.
Then $\{\rho_k\}$ converges to $\rho$ in trace norm implies
$$\lim\limits_{k\rightarrow\infty}\rho_k^{T_B}=\rho^{T_B}\eqno{(9)}$$ in
trace norm.

{\bf Proof} \quad Since $\rho_k$ converges to $\rho$ in trace norm
implies $\rho_k$ converges to $\rho$ entry-wise, thus $\rho_k^{T_B}$
converges to $\rho^{T_B}$ entry-wise as well. And it is obvious that
$\rho$ is separable, thus $\rho^{T_B}$ is also a state. This implies
that $\lim\limits_{k\rightarrow\infty}\rho_k^{T_B}=\rho^{T_B}$ with
respect to the trace norm since the trace norm topology coincide
with the weak-star topology on ${\mathcal S}(H_A\otimes
H_B)$.$\square$

{\bf The proof of Theorem 1} \quad We prove the inequality (6) firstly.
Denote by $\mathcal{S}_{s-p}$ the set of all separable pure states
in ${\mathcal S}(H_A\otimes H_B)$. If $\rho$ is separable, then it
admits a representation of the Bochner integral \cite{11}
$$\rho=\int_{{\mathcal S}_{s-p}}\varphi(\rho^A\otimes\rho^B)d\mu(\rho^A\otimes\rho^B),\eqno{(10)}$$
where $\mu$ is a Borel probability measure on $\mathcal{S}_{s-p}$,
$\rho^A\otimes\rho^B\in\mathcal{S}_{s-p}$
and  $\varphi:\mathcal{S}_{s-p}\rightarrow\mathcal{S}_{s-p}$ is a
measurable function. It immediately follows that
$$\rho_A=\int_{\mathcal{S}_{s-p}}\varphi(\rho^A\otimes\rho^B)^Ad\mu(\rho^A\otimes\rho^B),\eqno{(11)}$$
and
$$\rho_B=\int_{\mathcal{S}_{s-p}}\varphi(\rho^A\otimes\rho^B)^Bd\mu(\rho^A\otimes\rho^B),\eqno{(12)}$$
where $\varphi(\rho^A\otimes\rho^B)^A={\rm Tr}_B[\varphi(\rho^A\otimes\rho^B)]$,
$\varphi(\rho^A\otimes\rho^B)^B={\rm Tr}_A[\varphi(\rho^A\otimes\rho^B)]$.

Observe that
$$\begin{array}{rl}
&\rho-\rho_A\otimes\rho_B\\
=&\int_{\mathcal{S}_{s-p}}\varphi(\rho^A\otimes\rho^B)^A
\otimes\varphi(\rho^A\otimes\rho^B)^B d\mu(\rho^A\otimes\rho^B)\\
&-(\int_{\mathcal{S}_{s-p}}\varphi(\rho^A\otimes\rho^B)^Ad\mu(\rho^A\otimes\rho^B))\\
&\otimes(\int_{\mathcal{S}_{s-p}}\varphi(\rho^A\otimes\rho^B)^Bd\mu(\rho^A\otimes\rho^B))\\
=&\int_{\mathcal{S}_{s-p}}(\int_{\mathcal{S}_{s-p}}
\varphi(\rho^A\otimes\rho^B)^Ad\mu(\sigma^A\otimes\sigma^B))\\
&\otimes\varphi(\rho^A\otimes\rho^B)^Bd\mu(\rho^A\otimes\rho^B)\\
&-(\int_{\mathcal{S}_{s-p}}\varphi(\rho^A\otimes\rho^B)^Ad\mu(\rho^A\otimes\rho^B))\\
&\otimes(\int_{\mathcal{S}_{s-p}}\varphi(\rho^A\otimes\rho^B)^Bd\mu(\rho^A\otimes\rho^B))\\
=&\int_{\mathcal{S}_{s-p}}\int_{\mathcal{S}_{s-p}}\varphi(\rho^A\otimes\rho^B)^A
\otimes\varphi(\rho^A\otimes\rho^B)^B\\
&\cdot d\mu(\rho^A\otimes\rho^B)d\mu(\sigma^A\otimes\sigma^B)\\
&-\int_{\mathcal{S}_{s-p}}\int_{\mathcal{S}_{s-p}}\varphi(\sigma^A\otimes\sigma^B)^A\\
&\otimes\varphi(\rho^A\otimes\rho^B)^Bd\mu(\rho^A\otimes\rho^B)d\mu(\sigma^A\otimes\sigma^B)\\
=&\int_{\mathcal{S}_{s-p}}\int_{\mathcal{S}_{s-p}}(\varphi(\rho^A\otimes\rho^B)^A
\otimes\varphi(\rho^A\otimes\rho^B)^B\\
&-\varphi(\sigma^A\otimes\sigma^B)^A
\otimes\varphi(\rho^A\otimes\rho^B)^B)\\
&\cdot d\mu(\rho^A\otimes\rho^B)d\mu(\sigma^A\otimes\sigma^B)\\
=&\int_{\mathcal{S}_{s-p}}\int_{\mathcal{S}_{s-p}}(\varphi(\rho^A\otimes\rho^B)^A
-\varphi(\sigma^A\otimes\sigma^B)^A)\\
&\otimes\varphi(\rho^A\otimes\rho^B)^B
d\mu(\rho^A\otimes\rho^B)d\mu(\sigma^A\otimes\sigma^B)\\
=&\frac{1}{2}\int_{\mathcal{S}_{s-p}}\int_{\mathcal{S}_{s-p}}(\varphi(\rho^A\otimes\rho^B)^A
-\varphi(\sigma^A\otimes\sigma^B)^A)\\
&\otimes(\varphi(\rho^A\otimes\rho^B)^B-\varphi(\sigma^A\otimes\sigma^B)^B)\\
&\cdot d\mu(\rho^A\otimes\rho^B)d\mu(\sigma^A\otimes\sigma^B),\end{array}$$
where $\sigma^A\otimes\sigma^B\in\mathcal{S}_{s-p}$. We can arrive at
$$\begin{array}{rl}
&(\rho-\rho_A\otimes\rho_B)^R\\
=&\frac{1}{2}\int_{\mathcal{S}_{s-p}}\int_{\mathcal{S}_{s-p}}[(\varphi(\rho^A\otimes\rho^B)^A
-\varphi(\sigma^A\otimes\sigma^B)^A)\\
&\otimes(\varphi(\rho^A\otimes\rho^B)^B-\varphi(\sigma^A\otimes\sigma^B)^B)]^R\\
&\cdot d\mu(\rho^A\otimes\rho^B)d\mu(\sigma^A\otimes\sigma^B)\end{array}$$
with respect to the Hilbert-Schmidt norm since the realignment operation is continuous in
the Hilbert-Schmidt norm \cite{15}. It turns out that
$$\begin{array}{rl}
&\|(\rho-\rho_A\otimes\rho_B)^R\|_{\rm Tr}\\
\leq&\frac{1}{2}\int_{\mathcal{S}_{s-p}}\int_{\mathcal{S}_{s-p}}\|[(\varphi(\rho^A\otimes\rho^B)^A
-\varphi(\sigma^A\otimes\sigma^B)^A)\\
&\otimes(\varphi(\rho^A\otimes\rho^B)^B-\varphi(\sigma^A\otimes\sigma^B)^B)]^R\|_{\rm Tr}\\
&\cdot d\mu(\rho^A\otimes\rho^B)d\mu(\sigma^A\otimes\sigma^B).\end{array}$$
On the other hand, we let $\varphi(\sigma^A\otimes\sigma^B)^A=|x\rangle\langle x|$,
$\varphi(\sigma^A\otimes\sigma^B)^A=|y\rangle\langle y|$,
$\varphi(\rho^A\otimes\rho^B)^B=|f\rangle\langle f|$ and
$\varphi(\sigma^A\otimes\sigma^B)^B=|g\rangle\langle g|$, where $|x\rangle=(x_1$, $x_2$,
$\cdots$, $x_n$, $\cdots)^{T}$, $|y\rangle=(y_1$, $y_2$,
$\cdots$, $y_n$, $\cdots)^{T}\in H_A$, $|f\rangle=(f_1$, $f_2$,
$\cdots$, $f_n$, $\cdots)^{T}$ and $|g\rangle=(g_1$, $g_2$,
$\cdots$, $g_n$, $\cdots)^{T}\in H_B$. Then
$$\begin{array}{rl}&\|(\varphi(\sigma^A\otimes\sigma^B)^A
-\varphi(\sigma^A\otimes\sigma^B)^A)\\
&\otimes(\varphi(\rho^A\otimes\rho^B)^B
-\varphi(\sigma^A\otimes\sigma^B)^B)^{R}\|_{\rm Tr}\\
=&\|(|\varphi(\sigma^A\otimes\sigma^B)^A\rangle
-|\varphi(\sigma^A\otimes\sigma^B)^A\rangle)\\
&\cdot(\langle\varphi(\rho^A\otimes\rho^B)^B|
-\langle\varphi(\sigma^A\otimes\sigma^B)^B|)\|_{\rm Tr}\\
=&[\sum\limits_{i,j}(x_i\bar{x_j}-y_i\bar{y_j})
(\bar{x_i}x_j-\bar{y_i}y_j)]^{\frac{1}{2}}\\
&\cdot [\sum\limits_{i,j}(f_i\bar{f_j}-g_i\bar{g_j})
(\bar{f_i}f_j-\bar{g_i}g_j)]^{\frac{1}{2}}\\
=&[\sum\limits_{i,j}(|x_ix_j|^2+|y_iy_j|^2
-x_i\bar{x_j}\bar{y_i}y_j-\bar{x_i}x_jy_i\bar{y_j})]^{\frac{1}{2}}\\
&\cdot[\sum\limits_{i,j}(|f_if_j|^2+|g_ig_j|^2
-f_i\bar{f_j}\bar{g_i}g_j-\bar{f_i}f_jg_i\bar{g_j})]^{\frac{1}{2}}\\
=&(2-\sum\limits_{i,j}(x_i\bar{x_j}\bar{y_i}y_j
+\bar{x_i}x_jy_i\bar{y_j}))^{\frac{1}{2}}\\
&\cdot(2-\sum\limits_{i,j}(f_i\bar{f_j}\bar{g_i}g_j
+\bar{f_i}f_jg_i\bar{g_j}))^{\frac{1}{2}}\\
=&2[(1-{\rm Tr}(\varphi(\sigma^A\otimes\sigma^B)^A\varphi(\sigma^A\otimes\sigma^B)^A))\\
&\cdot (1-{\rm Tr}(\varphi(\rho^A\otimes\rho^B)^B
\varphi(\sigma^A\otimes\sigma^B)^B))]^{\frac{1}{2}}.\end{array}$$
Now, we have
$$\begin{array}{rl}&\|(\rho-\rho_A\otimes\rho_B)^{R}\|_{\rm Tr}\\
\leq&\int_{\mathcal{S}_{s-p}}\int_{\mathcal{S}_{s-p}}
[1-{\rm Tr}(\varphi(\sigma^A\otimes\sigma^B)^A
\varphi(\sigma^A\otimes\sigma^B)^A)]^{\frac{1}{2}}\\
&\cdot[1-{\rm Tr}(\varphi(\rho^A\otimes\rho^B)^B
\varphi(\sigma^A\otimes\sigma^B)^B)]^{\frac{1}{2}}\\
&\cdot d\mu(\rho^A\otimes\rho^B)d\mu(\sigma^A\otimes\sigma^B)\\
\leq&[\int_{\mathcal{S}_{s-p}}\int_{\mathcal{S}_{s-p}}
\|(1-{\rm Tr}(\varphi(\sigma^A\otimes\sigma^B)^A
\varphi(\sigma^A\otimes\sigma^B)^A))\\
&\cdot d\mu(\rho^A\otimes\rho^B)
d\mu(\sigma^A\otimes\sigma^B)]^{\frac{1}{2}}\\
&\cdot[\int_{\mathcal{S}_{s-p}}\int_{\mathcal{S}_{s-p}}
\|(1-{\rm Tr}(\varphi(\sigma^A\otimes\sigma^B)^B
\varphi(\sigma^A\otimes\sigma^B)^B))\\
&\cdot d\mu(\rho^A\otimes\rho^B)
 d\mu(\sigma^A\otimes\sigma^B)]^{\frac{1}{2}}\\
=&[(1-{\rm Tr}(\rho_A^2))(1-{\rm Tr}(\rho_B^2))]^{\frac{1}{2}}.\end{array}$$
(by Cauchy-Schwarz inequality we can obtain the second inequality.)

Now we begin to show the inequality (7). If $\rho$ is countably
separable, we let $\rho=\sum_ip_i\rho_i^{A}\otimes\rho_i^B$. Then,
by Lemma 2, we have
$$\begin{array}{rl}&\|(\rho-\rho_A\otimes\rho_B)^{T_B}\|_{\rm
Tr}\\
=&\|\frac{1}{2}\sum\limits_{i,j}p_ip_j(\rho_i^{A}-\rho_j^{
A})\otimes(\bar{\rho_i^{B}}
-\bar{\rho_j^{B}})\|_{\rm Tr}\\
\leq&\frac{1}{2}\sum\limits_{i,j}p_ip_j
\|(\rho_i^{A}-\rho_j^{A})\otimes(\bar{\rho_i^{B}}
-\bar{\rho_j^{B}})\|_{\rm Tr}\\
=&\frac{1}{2}\sum\limits_{i,j}p_ip_j
\|\rho_i^{A}-\rho_j^{A}\|_{\rm Tr}\|\bar{\rho_i^{B}}
-\bar{\rho_j^{B}}\|_{\rm Tr}\\
=&\frac{1}{2}\sum\limits_{i,j}p_ip_j
\|\rho_i^{A}-\rho_j^{A}\|_{\rm Tr}\|\rho_i^{B}
-\rho_j^{B}\|_{\rm Tr}
\end{array}$$ since $$\begin{array}{rl}
&\rho-\rho_A\otimes\rho_B\\
=&\sum\limits_ip_i\rho_i^{A}\otimes\rho_i^{B}
-(\sum_ip_i\rho_i^{A})\otimes(\sum_jp_j\rho_j^{B})\\
=&\sum\limits_{i,j}(p_j\rho_i^{A})\otimes(p_i\rho_i^{B})
-(\sum_ip_i\rho_i^{A})\otimes(\sum_jp_j\rho_j^{B})\\
=&\sum\limits_{i,j}[(p_j\rho_i^{A})\otimes(p_i\rho_i^{B})
-(p_i\rho_i^{A})\otimes(p_j\rho_j^{B})]\\
=&\sum\limits_{i,j}p_ip_j(\rho_i^{A}\otimes\rho_i^{B}
-\rho_i^{A}\otimes\rho_j^{B})\\
=&\frac{1}{2}\sum\limits_{i,j}p_ip_j(\rho_i^{A}-\rho_j^{A})
\otimes(\rho_i^{B}-\rho_j^{B}).\end{array}$$
Noticing that, ${\rm rank}(\rho_i^{A}-\rho_j^{A})\leq2$, ${\rm
Tr}(\rho_i^{A}-\rho_j^{A})=0$ and $(\rho_i^{
A}-\rho_j^{A})^\dagger=\rho_i^{A}-\rho_j^{A}$, we can conclude that
the eigenvalues of  $\rho_i^{A}-\rho_j^{A}$
are $\alpha$,
$-\alpha$, $\alpha\geq0$ which implies that the singular values of
$\rho_i^{A}-\rho_j^{A}$ are $\alpha$, $\alpha$. It follows from ${\rm
Tr}[(\rho_i^{A}-\rho_j^{A})^2]=2\alpha^2$ that
$\|\rho_i^{A}-\rho_j^{A}\|_{\rm Tr}=\sqrt{2{\rm
Tr}[(\rho_i^{A}-\rho_j^{A})^2]}=2\sqrt{1-{\rm
Tr}(\rho_i^{A}\rho_j^{A})}$. Similarly, we have $\|\rho_i^{
B}-\rho_j^{B}\|_{\rm Tr}=2\sqrt{1-{\rm Tr}(\rho_i^{
B}\rho_j^{B})}$. Thus, by Cauchy-Schwarz inequality, we arrive at
$$\|(\rho-\rho_A\otimes\rho_B)^{T_B}\|_{\rm
Tr}\leq2\sqrt{[1-{\rm Tr}(\rho_A^2)][1-{\rm Tr}(\rho_B^2)]}.$$

If $\rho$ is not countably separable, then there exists a sequence
of countably separable states $\{\sigma_n\}$ such that
$\lim\limits_{n\rightarrow}\sigma_n=\rho$ with respect to the trace
norm. It follows from Proposition 1 and Lemma 3 that,
$$\begin{array}{rl}&\|(\rho-\rho_A\otimes\rho_B)^{T_B}\|_{\rm Tr}\\
=&\lim\limits_{n\rightarrow\infty}
\|(\sigma_n-\sigma_{A(n)}\otimes\sigma_{B(n)})^{T_B}\|_{\rm Tr}\\
\leq&\lim\limits_{n\rightarrow\infty}2\sqrt{[1-{\rm
Tr}(\sigma_{A(n)}^2)][1-{\rm Tr}(\sigma_{B(n)}^2)]}\\
=&\sqrt{[1-{\rm
Tr}(\rho_A^2)][1-{\rm Tr}(\rho_B^2)]},\end{array}$$
 where
$\sigma_{A(n)}={\rm Tr}_B(\sigma_n)$ and $\sigma_{B(n)}={\rm Tr}_A(\sigma_n)$.$\Box$

 \if{\bf Remark}\quad
Let $\{\rho_n\}$ be a sequence in ${\mathcal S}_{sep}(H_A\otimes H_B)$.
In general, $\rho_n$ converges to
$\rho$ in trace norm doesn't imply $\{\rho_n^{R}\}$ converges to
$\rho^{R}$ in trace norm. For example, let
$\rho_n=a_n{\rm diag}(1,\frac{1}{2^{1+\frac{1}{n}}},\dots,\frac{1}{k^{1+\frac{1}{n}}},\dots)$, $n=1$, 2, $\dots$, where
$a_n=(\sum\limits_{k=1}^{+\infty}\frac{1}{k^{1+\frac{1}{n}}})^{-1}$.
It is clear that $\{\rho_n\}$ doesn't converge to $0$ in the trace norm.
However, $\|\rho_n^R\|_2^2={a_n^2}\sum\limits_k\frac{1}{k^{2+\frac{2}{n}}}
\leq{a_n^2} \sum\limits_k\frac{1}{k^{2}}\rightarrow0$, which reveals that $\rho_n^R$ converges to
$0$ in the Hilbert-Schmidt norm.\fi

\if Together with the Eqs.(2)-(3), we get the following result.

{\bf Proposition 2} \ \ \  Let $H_A$ and $H_B$ be complex separable Hilbert spaces with  $\dim H_A\otimes H_B\leq+\infty$. If
$\rho\in{\mathcal S}(H_A\otimes H_B)$ is separable, then
$$\|(\rho-\rho_A\otimes\rho_B)^{R}\|_{\rm Tr}\leq\sqrt{[1-{\rm
Tr}(\rho_A^2)][1-{\rm Tr}(\rho_B^2)]}\eqno{(13)}$$ and
$$\|(\rho-\rho_A\otimes\rho_B)^{T_B}\|_{\rm
Tr}\leq2\sqrt{[1-{\rm Tr}(\rho_A^2)][1-{\rm
Tr}(\rho_B^2)]}.\eqno{(14)}$$
Where $\rho_A={\rm Tr}_B(\rho)$
and $\rho_B={\rm Tr}_A(\rho)$.\fi

We assert that inequality (6) can
detect all states that can be recognized by the CCNR criterion. In
fact, if $\|\rho^R\|_{\rm Tr}>1$, then
$\|(\rho-\rho_A\otimes\rho_B)^R\|_{\rm Tr} \geq\|\rho^R\|_{\rm
Tr}-\|(\rho_A\otimes\rho_B)^R\|_{\rm Tr} =\|\rho^R\|_{\rm
Tr}-\||\rho_A\rangle\langle\rho_B|\|_{\rm Tr} =\|\rho^R\|_{\rm
Tr}-\|\rho_A\|_2\cdot\|\rho_B\|_2
>1-\|\rho_A\|_2\cdot\|\rho_B\|_2$. Therefore,
$1-\|\rho_A\|_2\cdot\|\rho_B\|_2\geq\sqrt{[1-{\rm
Tr}(\rho_A^2)][1-{\rm Tr}(\rho_B^2)]}$.
In what follows, we will show that the inequality (6) in Theorem 1 provides
a criterion that can detect some PPT entangled state $\rho$ with $\|\rho^R\|_{\rm Tr}\leq1$.

{\bf Example} \quad Let $H_A$ and $H_B$ be complex Hilbert spaces with
orthonormal bases $\{|0\rangle,|1\rangle$, $|2\rangle$, $\dots\}$
and $\{|0'\rangle, |1'\rangle$, $|2'\rangle$, $\dots\}$,
respectively. Let $0<a<1$ and
$$\begin{array}{rl}\tilde{\rho}
=&\frac{a}{8a+1}(|00'\rangle\langle 00'|+|01'\rangle\langle 01'|+|02'\rangle\langle02'|\\
&+|00'\rangle\langle11'|+|00'\rangle\langle22'|+|11'\rangle\langle00'|+|22'\rangle\langle00'|\\
&+|10'\rangle\langle10'|+|11'\rangle\langle11'|+|12'\rangle\langle12'|\\
&+|11'\rangle\langle22'|+|22'\rangle\langle11'|+|21'\rangle\langle21'|)\\
&+\frac{1+a}{2}(|20'\rangle\langle20'|+|22'\rangle\langle22'|)\\
&+\frac{\sqrt{1-a^2}}{2}(|20'\rangle\langle22'|+|22'\rangle\langle20'|).\end{array}$$
Write $$\tilde{\rho_\epsilon}=\epsilon\tilde{\rho}+(1-\epsilon)\frac{P_3}{9}, \quad
P_3=\sum\limits_{i,j=0}^2|i\rangle\langle i|\otimes|j'\rangle\langle j'|.$$
If $\dim H_A=\dim H_B=3$, it is obvious that
$$\tilde{\rho}=\hat{a}\left(\begin{array}{ccc|ccc|ccc}
a&0&0&0&a&0&0&0&a\\
0&a&0&0&0&0&0&0&0\\
0&0&a&0&0&0&0&0&0\\ \hline
0&0&0&a&0&0&0&0&0\\
a&0&0&0&a&0&0&0&a\\
0&0&0&0&0&a&0&0&0\\ \hline
0&0&0&0&0&0&\frac{1}{2}(1+a)&0&\frac{\sqrt{1-a^2}}{2}\\
0&0&0&0&0&0&0&a&0\\
a&0&0&0&a&0&\frac{\sqrt{1-a^2}}{2}&0&\frac{1}{2}(1+a)\end{array}\right)$$
(is a bound entangled state \cite{19}) and
$$\tilde{\rho_\epsilon}=\epsilon\rho+(1-\epsilon)\frac{I}{9},\quad \hat{a}=\frac{1}{8a+1}.$$
It is showed in \cite{13} that,
for $3\otimes3$ system, $\tilde{\rho_\epsilon}$ is entangled when $\epsilon\geq0.9955$ and $a=0.236$
applying the CCNR criterion. Using inequality (2), it is found that
$\tilde{\rho_\epsilon}$ is still entangled when $\epsilon=0.9939$ and $a=0.232$ \cite{14}.
It is straightforward that
$$\tilde{\rho_\epsilon}\ \mbox{\rm is entangled whenever}\ \epsilon\geq0.9939\ \mbox{\rm and}\ a=0.232.$$
Define
$$\sigma=\sum\limits_{i=3}^{+\infty}p_i|i\rangle\langle i|\otimes|i'\rangle\langle i'|,\quad
p_i\geq0,\ \sum\limits_{i=3}^{+\infty}p_i=1.$$ It is clear that
$\sigma$ is separable. Now we let
$$\tilde{\rho_{\epsilon,c}}=c\tilde{\rho_\epsilon}+(1-c)\sigma,\quad 0\leq c\leq 1,\eqno{(15)}$$
then $\|\tilde{\rho_{\epsilon,c}}^R\|_{\rm
Tr}=c\|\tilde{\rho_\epsilon}^R\|_{\rm Tr}+1-c$ since
$\|\sigma^R\|_{\rm Tr}=1$ and it is evident that
$\tilde{\rho_{\epsilon,c}}^{T_{A\setminus B}}\geq0$. On the other
hand, one can find that $\check{\rho_A}={\rm
Tr}_A(\tilde{\rho_{\epsilon,c}}) =c{\rm
Tr}_A(\tilde{\rho_\epsilon})+(1-c){\rm Tr}_A(\sigma)$ and
$\check{\rho_B}={\rm Tr}_B(\tilde{\rho_{\epsilon,c}}) =c{\rm
Tr}_B(\tilde{\rho_\epsilon})+(1-c){\rm Tr}_B(\sigma)$. Together with
the fact that the trace operation is completely bounded, we can
conclude that there exists some $0<c_0<1$,
$0.9939\leq\epsilon_0<0.9955$ and $0<\varepsilon<0.232$ such that
$\tilde{\rho_{\epsilon,c}}$ violates the inequality (6) whenever
$c>c_0$, $0.9939\leq\epsilon<\epsilon_0$ and $0.232-\varepsilon< a<0.232+\varepsilon$ while
$\|\tilde{\rho_{\epsilon,c}}^R\|_{\rm Tr}\leq1$ and
$\tilde{\rho_{\epsilon,c}}^{T_{A\setminus B}}\geq0$ whenever
$c>c_0$, $0.9939\leq\epsilon<\epsilon_0$ and $0.232-\varepsilon< a<0.232+\varepsilon$.

\section{Conclusions}

In conclusion, an entanglement criterion beyond the CCNR criterion for
infinite-dimensional systems is proposed: \if We
find that, similar as the finite-dimensional case, the reduced
density operator can also be derived via the partial trace
operation. Moreover, the partial trace operation is completely
bounded with respect to the trace norm topology on the set of all
states.\fi Based on the CCNR criterion for infinite-dimensional
systems, we highlighted the relation between separable states and
the reduced states via realignment operation or partial
transposition; It is
showed that the obtained inequality can detect more entangled states than the CCNR criterion.
It should be stressed here that the proof of our main result needs new tools which
is very different from the finite-dimensional case.

{\bf Acknowledgements.}\  This work is partially supported by
Natural Science Foundation of China (11171249,11101250) and Research
start-up fund for Doctors of Shanxi Datong University (2011-B-01).


\begin{thebibliography}{}
%
%
\bibitem{1} Nielsen M A, Chuang I L. Quantum Computatation and
Quantum Information. Cambridge: Cambridge University Press, 2000


\bibitem{2} Horodecki R, Horodecki P, Horodecki M, Horodecki K.
Quantum entanglement. Rev Modern Phys, 2009, 81,April-June

\bibitem{3} G\"{u}hne O, T\'{o}th G.
Entanglement detection. Phys Reports, 2009, 474: 1--75

\bibitem{4} Hou J C. A characterization of positive
linear maps and criteria for entangled quantum states.
J Phys A: Math Theor, 2010, 43, 385201

\bibitem{5} Hou J C, Qi X F.
Constructing entanglement witness for infinite-dimensional systems.
Phys Rev A, 2010, 81, 062351

\bibitem{6} Hou J C, Guo Y.
When different entanglement witnesses detect the same entangled states.
Phys Rev A, 2010, 82, 052301

\bibitem{7} Hou J C, Guo Y.
Constructing entanglement witnesses for states in infinite-dimensional bipartite quantum systems.
Int J Theor Phys, 2011, 50: 1245--1254

\bibitem{8} Qi X F, Hou J C.
Positive finite rank elementary operators and characterizing entanglement of states.
J Phys A: Math Theor, 2011, 44: 215305




\bibitem{Hou6} Qi X F, Hou J C. Characterization of optimal entanglement witnesses.
Phys. Rev. A, 2012, 85: 022334.

\bibitem{Hou7} Guo Y, Hou J C. Comment on ``Remarks on the structure of states of composite quantum systems
and envariance'' [Phys. Lett. A 355 (2006) 180].
Phys. Lett. A, 2011, 375: 1160--1162.

\bibitem{Hou8} Guo Y, Hou J C, Wang Y C.
Concurrence for infinite-dimensioanl quantum systems
arXiv:1203.3933v1(2012).

\bibitem{Hou9} Guo Y, Hou J C.
Detecting quantum correlations by means of local commuatativity,
arXiv:1107.0355v3(2011).

\bibitem{9} Horodecki M, Horodecki P, Horodecki R. Separability of
mixed states: necessary and sufficient conditions. Phys Lett A, 1996, 223: 1--8

\bibitem{10} Werner R F. Quantum states with Einstein-Posolsky-Rosen correlations
asmitting a hidden-variable model. Phys Rev A, 1989, 40, 4277

\bibitem{11} Holevo A S, Shirokov M E, Werner R F.
Separability and entanglement-breaking in infinite-dimensions.
Russian Math Surveys, 2005, 60: N2

\bibitem{12} Rudolph O. Computable cross-norm criterion for
separability. Lett Math Phys, 2004, 70: 57--64

\bibitem{13} Chen K, Wu L A. A matrix realignment method
for recognizing entanglement. Quant Inf Comput, 2003, 3: 193--202

\bibitem{14} Zhang C J, Zhang Y S, Zhang S, Guo G C. Entanglement
detection beyond the computable cross-norm or realignment criterion.
Phys Rev A, 2008, 77: 060301(R)

\bibitem{15} Guo Y, Hou J C. The CCNR criterion of separability for states in
infinite-dimensional quantum systems. arXiv: 1009.0116v1

\bibitem{16} Guo Y, Qi X F, Hou J C. Sufficient and necessary
conditions of separability for bipartite pure states in
infinite-dimensional systems. Chinese Science Bull, 2011, 56(9): 840--846

\bibitem{17} Aniello P, Lupo C. A class of inequalities inducing new separability
criterion for bipartite quantum systems.
J Phys A: Math Theor, 2008, 41: 355303

\bibitem{18} Zhu S, Ma Z H. Topologies on quantum states.
Phys Lett A, 2010, 374: 1336--1341

\bibitem{19} Horodecki P. Separability criterion and inseparable mixed states with
positive partial transposition. Phys Lett A, 1997, 232, 333--339










\end{thebibliography}
\end{document}